\documentclass[10pt]{article}

\usepackage{amsmath}
\usepackage{amssymb}

\usepackage{graphicx}

\usepackage{cite}

\usepackage{color,ulem} 

\usepackage{setspace} 
\usepackage[labelfont=bf,labelsep=period,justification=raggedright]{caption}

\makeatletter
\renewcommand{\@biblabel}[1]{\quad#1.}
\makeatother

\date{}

\usepackage{subfigure}

\begin{document}

\begin{flushleft}
{\Large
\textbf{Quantifying the dynamics of coupled networks of switches and oscillators}
}
\\
Matthew R. Francis$^{1}$, 
Elana J. Fertig$^{2,\ast}$, 
\\
\bf{1} Physics Department, Randolph-Macon College, Ashland, VA, USA
\\
\bf{2} Department of Oncology and Division of Oncology Biostatistics and Bioinformatics, Sidney Kimmel Comprehensive Cancer Center, School of Medicine, Johns Hopkins University, Baltimore, MD, USA
\\
$\ast$ E-mail: Corresponding ejfertig@jhmi.edu
\end{flushleft}

\section*{Abstract}
Complex network dynamics have been analyzed with models of systems of coupled switches or systems of coupled oscillators.  However, many complex systems are composed of components with diverse dynamics whose interactions drive the system's evolution.  We, therefore, introduce a new modeling framework that describes the dynamics of networks composed of both oscillators and switches.  Both oscillator synchronization and switch stability are preserved in these heterogeneous, coupled networks.  Furthermore, this model recapitulates the qualitative dynamics for the yeast cell cycle consistent with the hypothesized dynamics resulting from decomposition of the regulatory network into dynamic motifs.  Introducing feedback into the cell-cycle network induces qualitative dynamics analogous to limitless replicative potential that is a hallmark of cancer.   As a result, the proposed model of switch and oscillator coupling provides the ability to incorporate mechanisms that underlie the synchronized stimulus response ubiquitous in biochemical systems.


\section*{Introduction}

\par The dynamics in systems ranging from intercellular gene regulation to organogenesis are driven by complex interactions (represented as edges) in subcomponents (represented as nodes) in networks.  If the structure of these networks is known, network-wide models of coupled systems have been applied to predict their qualitative dynamics.  For example, models of coupled switches based upon Glass networks \cite{Glass1973} have been applied to model systems such as neuronal synapses \cite{Hopfield1982} and gene regulatory networks \cite{Shmulevich2002b}.  Similarly, models of coupled oscillators along networks based upon the Kuramoto model \cite{Kuramoto1975} have been used to model synchronization of oscillators in diverse systems reviewed in \cite{Strogatz2003}.  In biochemical systems, \textit{in vivo} oscillator synchronization has been observed in synthetic oscillatory fluorescent bacteria \cite{Garcia-Ojalvo2004,Danino2010}, yeast gene regulatory networks \cite{Tu2005,Klevecz2008}, and human cell fate decisions \cite{Huang2005}.  Such spontaneous synchronization has also been attributed to the development of the mammalian cardiac pacemaker cells (reviewed in \cite{Peskin1975}) and cortical systems (reviewed in \cite{Breakspear2010}) including notably the circadian pacemaker (e.g., \cite{Strogatz1987}). More recently,  these network models have been found to be insufficient to model more complex dynamics in neuronal information transfer \cite{Fox2001,Cumin2007,Wang2008,Breakspear2010,Perez2011} and cardiac arrhythmias \cite{Kunysz1997,Chang2000,Sato2009,Zahanich2011}.  These limitations extend to physical systems, such as the coupled lasers studied in \cite{TianaAlsina2011}.  Therefore, numerous studies have modified these network models to account for evolving networks \cite{Cumin2007,Mondal2008,Sorrentino2008,Jiang2009,Volman2010, Adhikari2011,Gomez-Gardenes2011}, dynamic frequencies \cite{Cumin2007,Martens2009,Taylor2010}, or phase delays \cite{Wang2008,Ermentrout2009,Wang2009,Wang2011}.  However, these mathematical modifications typically do not encode the mechanism underlying the limitations in the Kuramoto and Glass network models.

\par We hypothesize that the observed limitations in the standard Kuramoto and Glass models arise from their exclusion of coupling components with qualitatively different dynamics.  Several studies have inferred that biochemical systems contain ``network motifs" with both oscillatory and switch-like dynamics \cite{Alon2007,Novak2008}.  The dynamics of these motifs are inferred from the topology of subgraphs in the networks of these systems.  Their structures are statistically overrepresented in biochemical networks \cite{Milo2002,Shen-Orr2002} such as intracellular regulatory networks \cite{Rao2001}, implicating evolutionary preservation (and thus utility) of these network motifs \cite{Prill2005}.  The dynamics of these motifs have been used to model yeast cell cycle regulation  \cite{Csikasz-Nagy2008} and have been further confirmed in synthetic, designed biochemical circuits (reviewed in \cite{Purcell2010}).  Because these heterogeneous network motifs are all identified as components within a single biochemical network, their interactions must drive the global dynamics of the network \cite{Taylor2011}.  Previously, \cite{Kim2008} have shown that coupling small sets of heterogeneous network motifs ensures the robustness of motif dynamics and \cite{Taylor2011} have shown that coupling networks changes their dynamics in isolation.   However, the network-level dynamics that result from coupling oscillatory and switch-like components have not been studied comprehensively.

\par  In this paper, we develop a theoretical framework to quantify the network-wide dynamics resulting from coupling switches and oscillators.  This model is based upon introducing cross-coupling between the Kuramoto and Glass models, due to their wide success in modeling the dynamics in networks of oscillators and networks of switches, respectively.  Simulations with the proposed model across state-space in an all-to-all network yields four operational states: (1) switches remain ``on'' and oscillators synchronize, (2) switches are ``off'' and oscillators freeze, (3) switches fluctuate in sync with oscillators, and (4) switches fluctuate transitionally until oscillators freeze.  Further application of our model to the network motifs identified in yeast in \cite{Tyson2003} recapitulates the qualitative dynamics of the system observed in that study.  However, a simple rewiring of this cell-cycle network that introduces feedback causes a cancer-like sustained re-activation of the cell cycle machinery without regard for external signal growth signals.  These dynamics suggest that modeling cross-motif coupling may predict critical processes in the dynamics of biochemical networks with minimal parameterization.

\subsection*{The Kuramoto model of coupled oscillators}

\par Quantitative studies of coupled oscillators often apply the Kuramoto model of $M$
oscillators coupled in an all-to-all network.  In this model, the change in time $\dot{\theta}_i$ of the 
phase of the $i^{\mbox{th}}$ oscillator, $\theta_{i}$, is governed by
\begin{equation}
\dot{\theta_{i}} = \hat{\omega}_{i} + \frac{\kappa_{\theta,\theta}}{M} \sum_{j=1}^{M}\sin\left(\theta_{j} - \theta_{i} \right),
\label{eq:Kuramoto}
\end{equation}
where $\hat{\omega}_{i}$ is the natural frequency of the $i^{\mbox{th}}$
oscillator and $\kappa_{\theta,\theta} \ge 0$ is the coupling strength
of the oscillators \cite{Kuramoto1975}.  
Typically, the $\hat{\omega}_i$ values are drawn from a normal distribution centered at 0 with variance $\sigma_\omega$.

\par In the Kuramoto model, the phases of the oscillators will synchronize if $\kappa_{\theta,\theta}$
is above a threshold coupling strength $\hat{\kappa}_{\theta,\theta}$.  Such synchronization is 
quantified with the mean field of the oscillators as 
\begin{equation}
r_\theta e^{i\psi} = \frac{1}{M} \sum_{j=1}^{M} e^{i\theta_{j}}. \label{eq:rOscMean}
\end{equation}
 Here $\psi$ is the average phase of the oscillators and the coherence
$r_\theta$ represents the spread of the oscillators from that average
phase.  Based upon eq.~(\ref{eq:rOscMean}), $r_\theta=1$ if each $\theta_{i} = \psi$ and
$r_\theta=0$ if the values of $\theta_{i}$ are distributed uniformly between 
$[0, 2\pi)$ \cite{Strogatz2000}. 
  
\subsection*{Glass networks of coupled switches}

\par Coupled sets of $N$ switches, which adopt one of a set of binary states, are modeled with Glass networks \cite{Glass1973}.  These models describe the evolution of the $i^{\mbox{th}}$ switch ($\tilde{x}_{i}$) as follows
\begin{equation}
\dot{x}_{i} = -x_{i} + F_{i}\left(\tilde{x}_{1},\tilde{x}_{2},\ldots,\tilde{x}_{n}\right)\mbox{, and}
\label{eq:GlassCont}
\end{equation}
\begin{equation}
\tilde{x}_{i} = 0 \mbox{ if } x_{i}<0 \mbox{; }1\mbox{ otherwise} ,
\label{eq:booleanx}
\end{equation}
where $\dot{x}_{i}$ represents the change in time of the value of each $x_{i}$, which are unobservable continuous variables that control the time of switching between observable, discrete states in $\tilde{x}_{i}$.  In this model, $F_{i}$ describes the change in state of the $i^{\mbox{th}}$ switch due to the coupling with the other $N$ switches in the network \cite{Glass1973}.  In specified network structures and functions $F_{i}$, such Glass networks can exhibit complex dynamics, including periodic and aperiodic orbits (e.g., \cite{Edwards2000}).

\par One type of Glass network, called a Hopfield network \cite{Hopfield1982}, has dynamics applicable to the smooth-decay of signal in biochemical switches \cite{Hopfield1982}.  The Hopfield model lets
\begin{equation}
F_{i} = \kappa_{x,x} \sum_{j=1}^{N}w_{ij}\tilde{x}_{j} - \tau_{i},
\label{eq:Hopfield}
\end{equation}
where $w_{ij}$ takes values between $-1$ and $1$ representing the relative strength of the connection between switches $i$ and $j$, $\kappa_{x,x}$ is the magnitude of coupling strengths, and $\tau_{i}$ the threshold for switch activation.  Similar to the Kuramoto model, sets of the switches will synchronize for $\kappa_{x,x}$ above a threshold $\hat{\kappa}_{x,x}$ in appropriate network topologies.  

\section*{Results}

\subsection*{Network model of coupled oscillators and switches}

\par By combining the established models for switches and oscillators, we model the dynamics of the heterogeneous system of coupled switches and oscillators in systems including biochemical networks with the following set of equations:
\begin{eqnarray}
\dot{x}_{i} & = & -x_{i} +  G_{i}\left(\tilde{x}_{1},\tilde{x}_{2},\ldots,\tilde{x}_{N}, \theta_{1},\theta_{2},\ldots,\theta_{M}\right) \label{eq:genSwitch} \\
\dot{\theta}_{l} & = & \omega_{l}\left(\tilde{x}_{1},\tilde{x}_{2},\ldots,\tilde{x}_{N}\right) \label{eq:genOscillator} \\
\nonumber & + &H_{l}\left(\tilde{x}_{1},\tilde{x}_{2},\ldots,\tilde{x}_{N}, \theta_{1},\theta_{2},\ldots,\theta_{M}\right) \label{eq:genOmega}.
\end{eqnarray}
Here, eq.~(\ref{eq:genSwitch}) is analogous to the Glass network in eq. (\ref{eq:GlassCont}) and $\tilde{x}_{i}$ is defined according to eq. (\ref{eq:booleanx}).

\par In this study, we explore a case of the switch-oscillator model in eqs.~(\ref{eq:genSwitch}) and (\ref{eq:genOscillator}) which contains an all-to-all network that couples the Kuramoto model, eq.~(\ref{eq:Kuramoto}), and Hopfield network, eqs.~(\ref{eq:GlassCont}) - (\ref{eq:Hopfield}), as follows
\begin{eqnarray}
\dot{x}_{i} & = & -x_{i} + \frac{\kappa_{x,x}}{N}\sum_{j\neq.~i}^{N}\tilde{x}_{j} + \frac{\kappa_{x,\theta}}{M}\sum_{k=1}^{M}\tilde{\theta}_{k} - \tau_{i}, \label{eq:switch} \\
\dot{\theta}_{l} & = & \omega_{l} + \frac{\kappa_{\theta,\theta}}{M} \sum_{k=1}^{M}\sin\left(\theta_{k} - \theta_{l} \right), \label{eq:osc} \\
\dot{\omega}_{l} & = &  \frac{\kappa_{\theta, x} }{N}\sum_{j=1}^{N}(\tilde{x}_{j} \hat{\omega}_l - \omega_l),\label{eq:omega}
\end{eqnarray}
where $\kappa_{x,\theta}$ and $\kappa_{\theta,x}$ are cross-component coupling strengths.  In eq.~(\ref{eq:omega}), $\omega_{l}$ is the time-varying frequency of the $l^{\mbox{th}}$ oscillator resulting from switch coupling, with initial values $\omega_{l} (t = 0) = \hat{\omega}_{l}$, and 
\begin{equation}
\tilde{\theta}_{l} = \left \{ \begin{array}{ll}
	 1 & \mbox{ if } 0 \le \theta_{l}<\pi \\
	 0 & \mbox{ otherwise} \\ \end{array} \right. . \label{eq:binOscillator}
\end{equation}
 In this system, zero values of the cross-coupling parameters $\kappa_{x,\theta}$ and $\kappa_{\theta,x}$ cause the model to reduce to the standard uncoupled Kuramoto and Hopfield models.  Similar decoupling of the models occurs if the switch and oscillator systems are at vastly different timescales, determined by the $\tau_{i}$ and $\hat{\omega}_{l}$ parameters, respectively.  The transformation in eq.~(\ref{eq:binOscillator}) facilitates comparable switch-like dynamics in the oscillators when they interact with switches in eq.~(\ref{eq:switch}).   Nonzero switch-oscillator ($\kappa_{x,\theta}$) interactions will cause an oscillator in the ``up'' part ($\tilde{\theta}_{l} = 1$) of its cycle to feed energy into the switch in question, nudging it towards the ``on'' state if off or delaying its decay if already on.  Similarly, an ``on'' switch with a nonzero oscillator-switch interaction ($\kappa_{\theta,x}$) will feed energy into the oscillators causing them to cycle at their natural frequency if coupled to that switch.

\par By thus incorporating coupling between switches and oscillators within the framework established by the standard Kuramoto and Hopfield
models, the dynamics of our model in eqs.~(\ref{eq:switch}) -
(\ref{eq:omega}) can be analyzed within the framework of these well
established models.  Similar to analysis of the Kuramoto model and Glass
network, we summarize the dynamics of our system using order parameters.  For oscillators, we utilize the order parameter defined in eq.~(\ref{eq:rOscMean}).  We introduce a new
order parameter 
\begin{equation}
r_\omega(t) = \frac{1}{M}\sum_{\mu} \frac{\omega_\mu(t)}{\hat{\omega}_\mu}
\end{equation}
that tracks how closely each individual oscillator's
frequency $\omega_\mu$ corresponds to the natural frequency
$\hat{\omega}_\mu$.  Analogously, we measure the fraction of switches
that are in the ``on'' position at a given time using a
switch-switch order parameter defined by
\begin{equation}
r_x(t) = \frac{1}{N} \sum_{j} \tilde{x}_j(t).
\end{equation}
Both of these functions will have a maximum of 1 when all
switches are on, and minimum 0 if all switches are off.  

\subsection*{Simulation results in all-to-all networks}

\par
\begin{figure}[ht]
\begin{center}
\subfigure[]{
\includegraphics[width=0.45\textwidth]{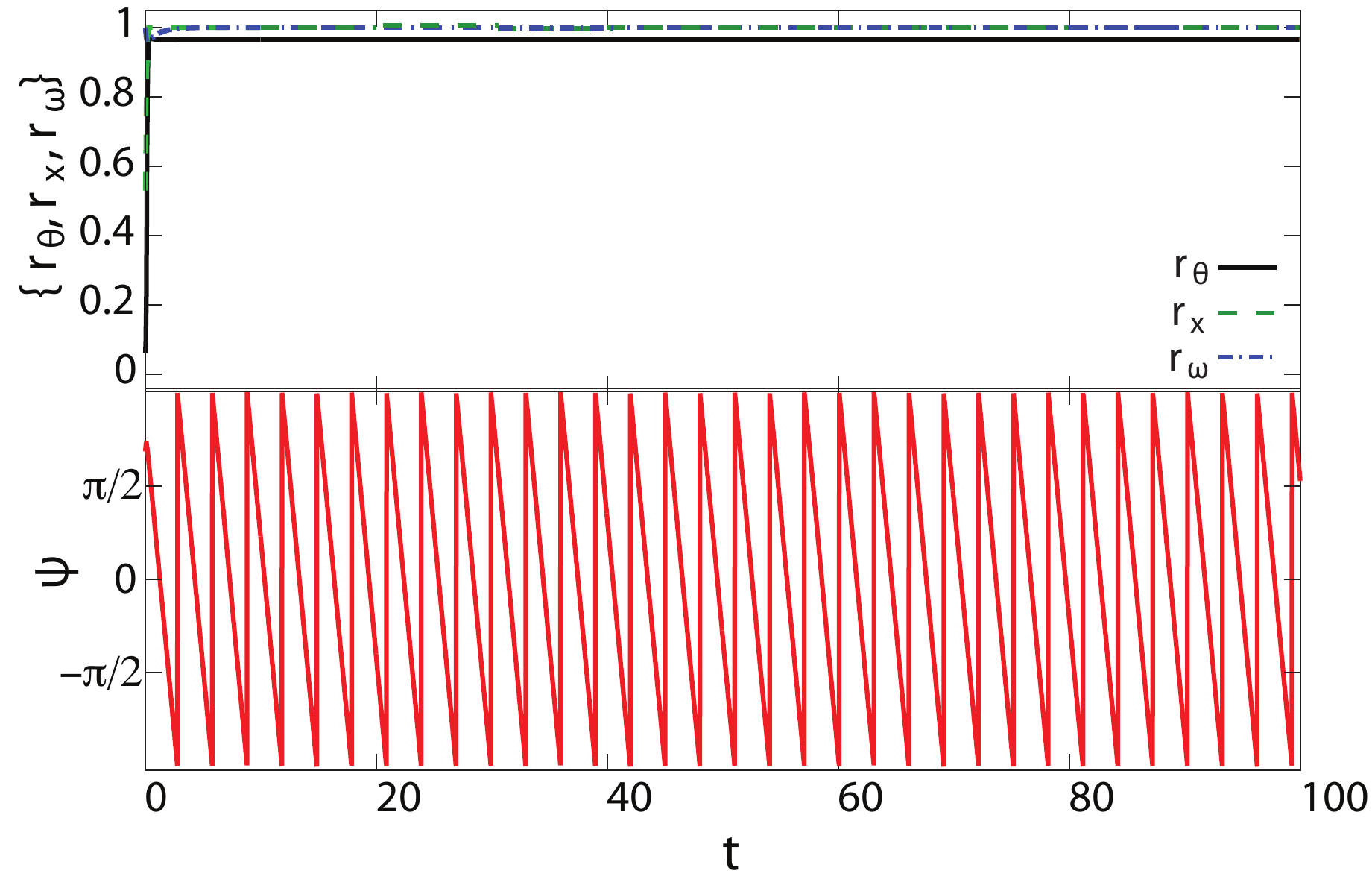}
\label{fig:coupledDynamicsOn}
}
\subfigure[]{
\includegraphics[width=0.45\textwidth]{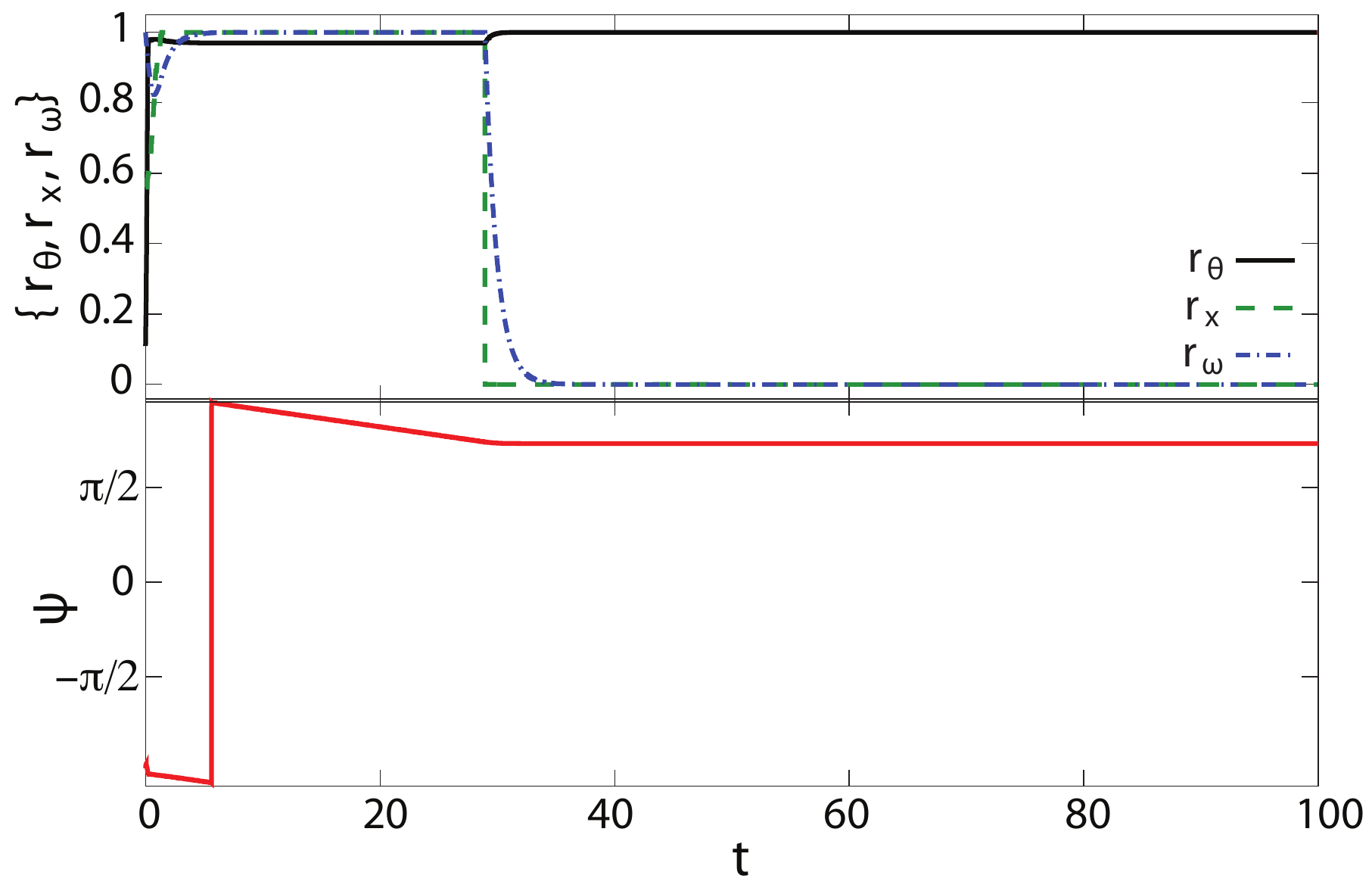}
\label{fig:coupledDynamicsOff}
}
\subfigure[]{
\includegraphics[width=0.45\textwidth]{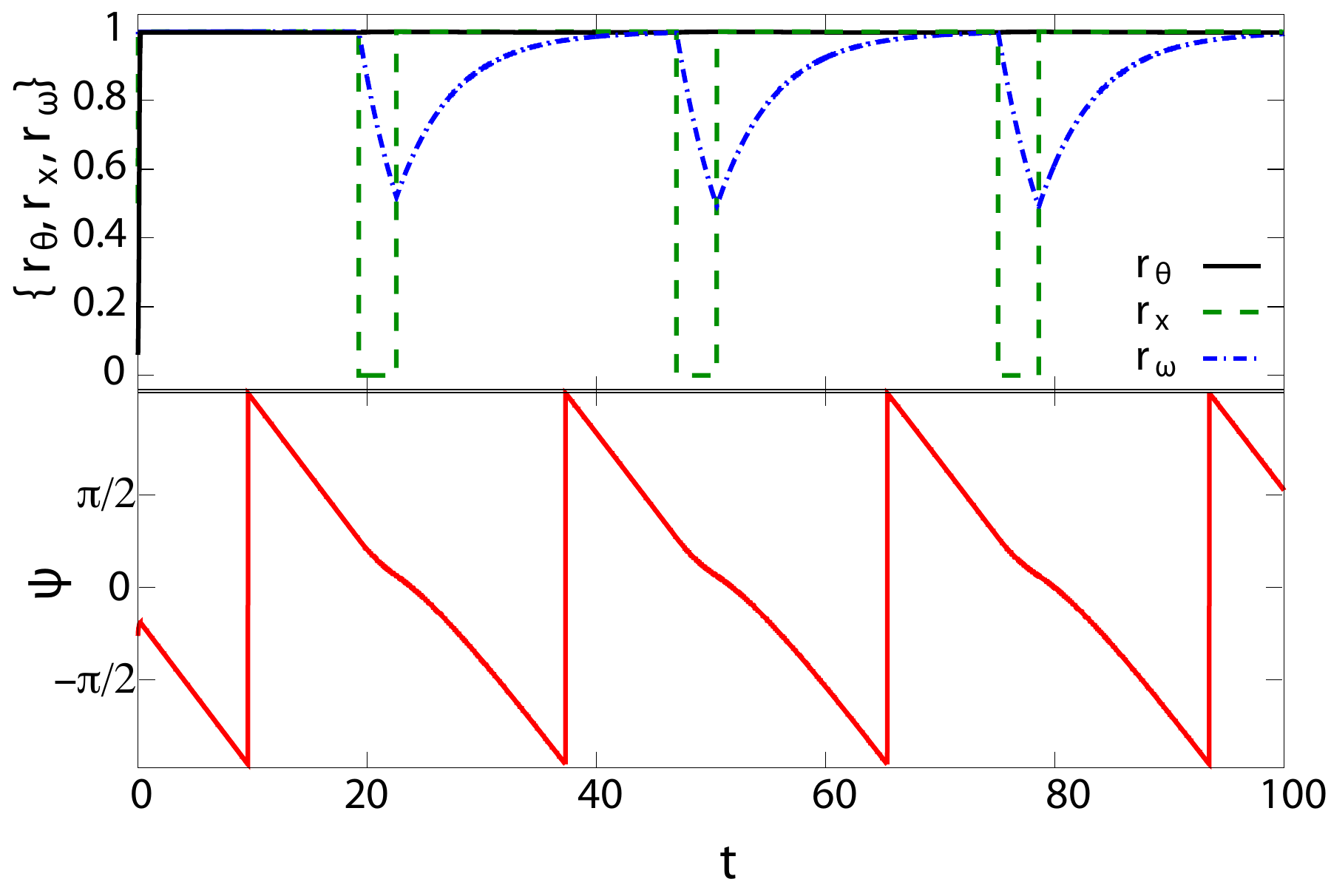}
\label{fig:coupledDynamicsOsc}
}
\subfigure[]{
\includegraphics[width=0.45\textwidth]{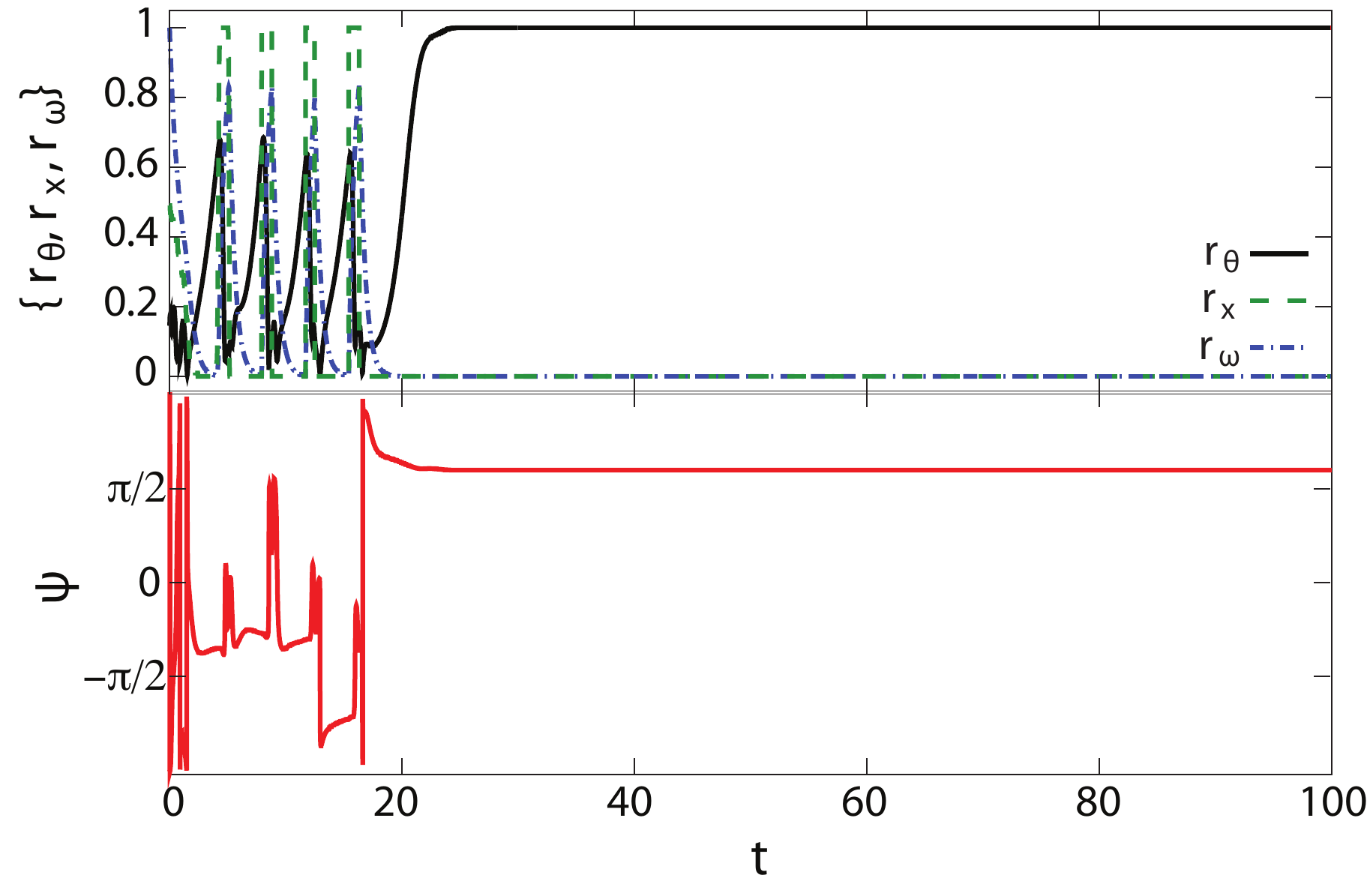}
\label{fig:coupledDynamicsOscNoSyncSwitchOsc}
}
\end{center}
\caption{{\bf Summary of the qualitative dynamics of the heterogeneous network model of eqs. (\ref{eq:switch}) -
(\ref{eq:omega}).} In all figures, top-panel shows temporal evolution of the mean field statistics ($r_{\theta}$ black, solid; $r_{x}$ green, dashed; and $r_{\omega}$ blue, dash-dotted) and the bottom-panel shows the evolution of the mean phase $\psi$ (red, solid). (a) Oscillators synchronize and switches stay ``on" ($\kappa_{x,x} = 11$, $\kappa_{x,\theta}=1.5$, 
$\kappa_{\theta,x}=1$, $\kappa_{\theta,\theta}=40$, and $\sigma_{\omega}=10$), (b) oscillators freeze (as evidenced by unchanging $\psi$) and switches stay ``off" ($\kappa_{x,x} = 1$, $\kappa_{x,\theta}=1.5$, $\kappa_{\theta,x}=1$, $\kappa_{\theta,\theta}=40$, and $\sigma_{\omega}=10$), (c) oscillators synchronize and switches oscillate ($\kappa_{x,x} = 1$, $\kappa_{x,\theta}=160$, $\kappa_{\theta,x}=0.2$, $\kappa_{\theta,\theta}=42$, and $\sigma_{\omega}=3$), and (d) transitory oscillations in oscillators and switches ($\kappa_{x,x} = 0.1$, $\kappa_{x,\theta}=1.4$, $\kappa_{\theta,x}=2$, $\kappa_{\theta,\theta}=1.8$, and $\sigma_{\omega}=10$).}\label{fig:coupledDynamics}
\end{figure}

  We first explore the qualitative dynamics of the heterogeneous system through numerical simulations in all-to-all networks.  We limited these simulations to all-to-all networks, because of the ability of this network topology to describe the qualitative dynamics from the Kuramoto model. These simulations explore the majority of parameter space defined by $\kappa_{x,x}$, $\kappa_{x,\theta}$, $\kappa_{\theta,\theta}$, $\kappa_{\theta,x}$, and $\sigma_{\omega}$.  Specifically, we select $\kappa_{x,x} = 1 < \hat{\kappa}_{x,x}$ to ensure that switches are able to turn off without appropriate stimulation from the oscillators.  We consider the effects of switches on oscillators for values of $\kappa_{\theta,\theta}$ both above and below the Kuramoto threshold $\hat{\kappa}_{\theta,\theta}$.  Figure~\ref{fig:coupledDynamics} plots the time-dependent order parameters observed in the four qualitative states observed in simulations of the coupled model eqs. (\ref{eq:switch}) - (\ref{eq:omega}) that are reflective of the qualitative dynamics observed in simulations with these parameter values.  Supplemental videos S1-S4 further summarize the results of these simulations.  We note that these four states were the only qualitative states observed for our coupled model in all-to-all networks simulated according to the description in the Methods section. Because $\tau$, $\kappa_{x,\theta}$, and $\sigma_{\omega}$ all control the relative timing of switches and oscillators, their values were selected in these simulations to optimize visualization in the supplemental videos.   When exploring the effect of timing on the system dynamics, we hold $\tau$ and $\kappa_{x,\theta}$ fixed while varying $\sigma_{\omega}$. Figure~\ref{fig:modelBehaviorSens} shows the probability of observing the states in Figures~\ref{fig:coupledDynamicsOn}-\ref{fig:coupledDynamicsOsc} in 100 simulations of all-to-all systems containing 100 switches and oscillators as a function of $\kappa_{\theta,x}$ and $\sigma_{\omega}$.  Because of their common control of system timing, we would obtain comparable distributions when varying either $\tau$ or $\kappa_{x,\theta}$ instead of $\sigma_{\omega}$.

\subsubsection*{The coupled system preserves synchronization in both oscillators and switches.}

\par Figure~\ref{fig:coupledDynamicsOn} shows a state of the model in which the switches are all in the ``on" state and oscillators are synchronized ($r_x$ near 1, $r_\theta$ near 1, and $\psi$ oscillating between $[0, 2\pi)$ periodically). While such synchronization is observed in the uncoupled Hopfield and Kuramoto models, the oscillator-switch cross coupling extends the region of parameter space over which this synchronization occurs.  Specifically, modest values of $\kappa_{x,\theta}$ can induce sustained switch activity for parameter values of $\kappa_{x,x}$ in which switches would decay in the uncoupled system.  Furthermore, this switch synchronization will occur for all values of $\kappa_{x,x}$ in which synchronization occurs in the uncoupled Hopfield model (i.e., all $\kappa_{x,x}$ larger than a threshold value $\hat{\kappa}_{x,x}$) because the oscillators only contribute positively to the derivative in eq.~(\ref{eq:switch}) in our model.  On the other hand, no value of $\kappa_{\theta,x}$ will cause oscillator phases to synchronize if $\kappa_{\theta,\theta}$ is below the critical coupling parameter for the pure Kuramoto model ($\hat{\kappa}_{\theta,\theta}$).  However, there are parameter regimes in which this synchronization occurs stochastically, depending on the initial values selected for $x_{i}$, $\theta_{j}$, and $\hat{\omega}_{j}$ (Figure~\ref{fig:modelBehaviorSensOn}).  In these cases, the average decrease in oscillator natural frequencies caused by decreasing $\kappa_{\theta,x}$ or $\sigma_{\omega}$ will increase the effective period of oscillators, thereby increasing the probability of switches being locked in the ``on'' state and oscillator synchronization in the heterogeneous system.

\subsubsection*{Coupling switches to unsynchronized oscillators can freeze network-wide dynamics.}

\par Figure~\ref{fig:coupledDynamicsOff} depicts a model state in which switches are all ``off" ($r_x$ near zero) and oscillators ``freeze": each $\theta_{j}\left(t\right) = \psi\left(t\right) = \Psi$ for some constant values $\Psi$ for all $t$ beyond the preliminary freezing time $t_{f}$.  While the decaying switches are observed in an uncoupled Hopfield model, the freezing oscillators cannot be simulated in the uncoupled Kuramoto model.  Such oscillator freezing will occur whenever the oscillators decay to the ``off" state by virtue of the coupling of the oscillators to switches through the $\omega_j$ in eq.~(\ref{eq:omega}). Specifically, this frozen state can occur whenever $\kappa_{x,x} < \hat{\kappa}_{x,x}$ depending on the values of $\tilde{x}_{i}$, $\theta_{j}$, and $\hat{\omega}_{j}$.  However, the probability of selecting these initial states is decreased when the heterogeneity of the oscillators increases through incomplete synchronization ($r_{\theta}(t) < 1$) or increased $\sigma_{\omega}$ (Figure~\ref{fig:modelBehaviorSensOff}).  In these cases, a single oscillator in the ``up" phase ($\tilde{\theta} = 1$) can contribute positively to the switch states, forcing the system out of this frozen state.  The probability of obtaining this frozen model state further depends on the relative timing of switch decay and oscillator freezing.  Specifically, the probability of obtaining the frozen state decreases with the average oscillator frequency, determined predominantly by the parameter $\kappa_{\theta,x}$ (Figure~\ref{fig:modelBehaviorSensOff}).

\subsubsection*{Coupling switches to synchronized oscillators can induce synchronized oscillations in switches.}

\par
\begin{figure}[ht]
\begin{center}
\subfigure[]{
\includegraphics[width=0.3\textwidth]{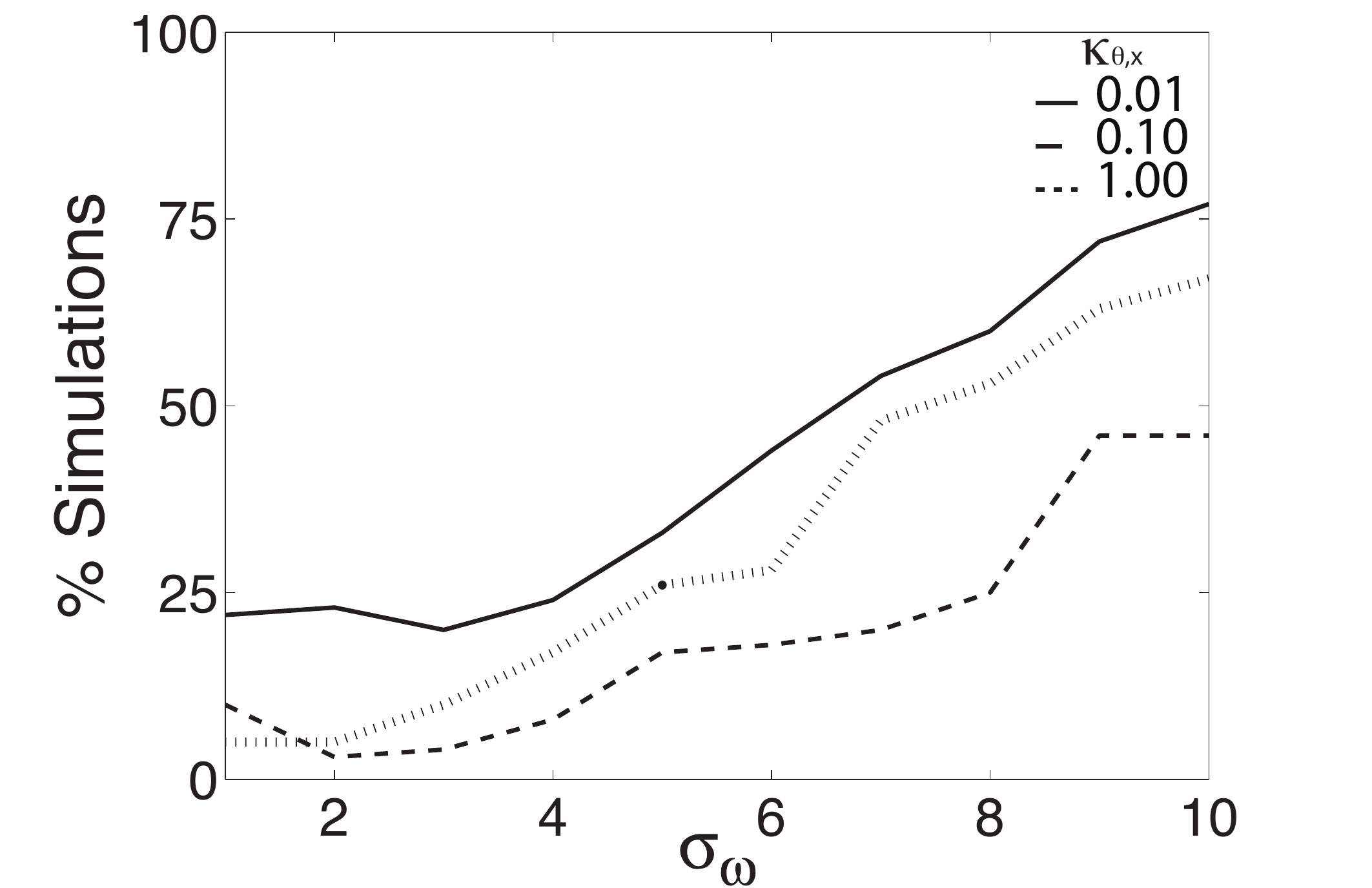}
\label{fig:modelBehaviorSensOn}
}
\subfigure[]{
\includegraphics[width=0.3\textwidth]{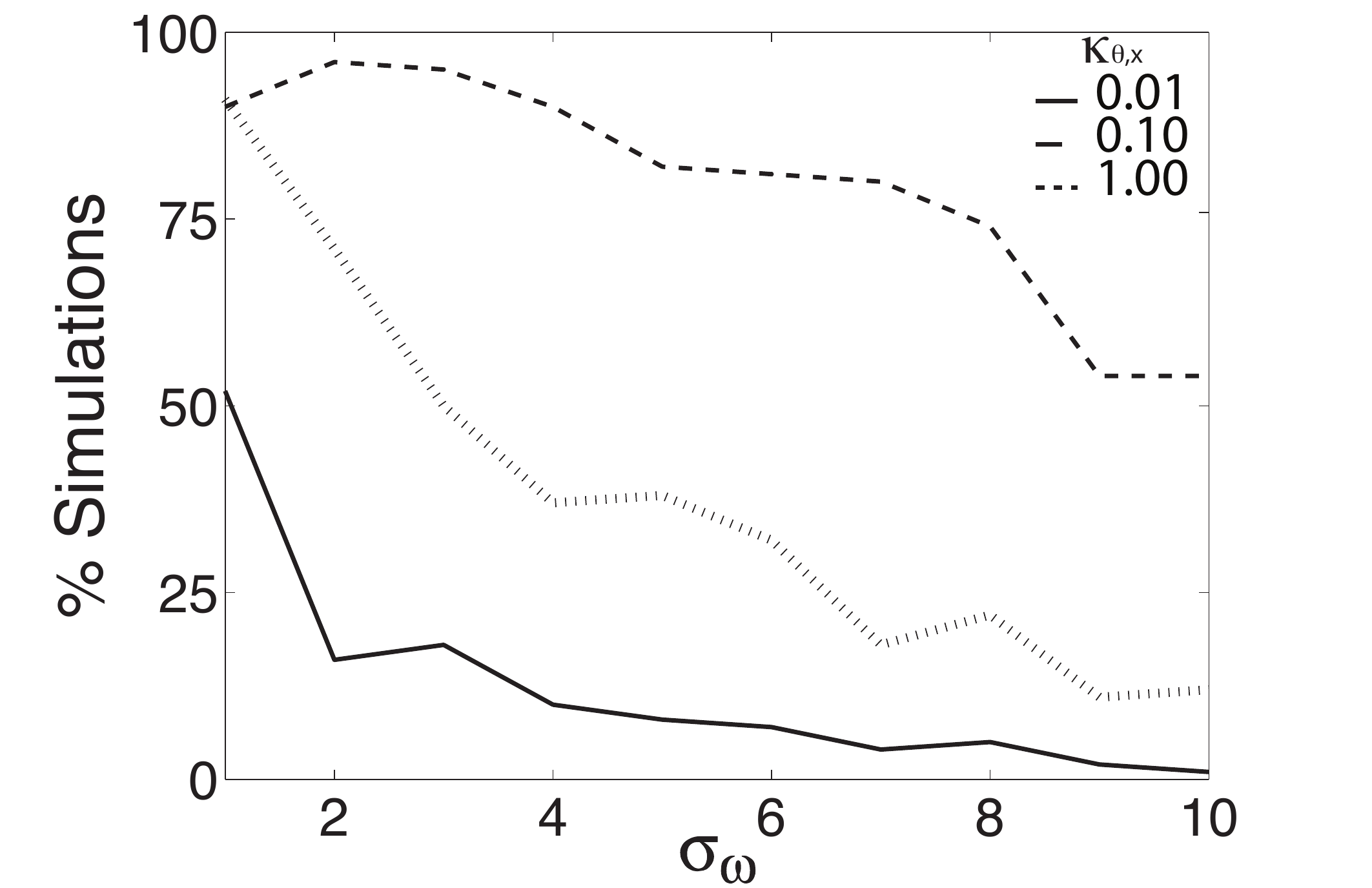}
\label{fig:modelBehaviorSensOff}
}
\subfigure[]{
\includegraphics[width=0.3\textwidth]{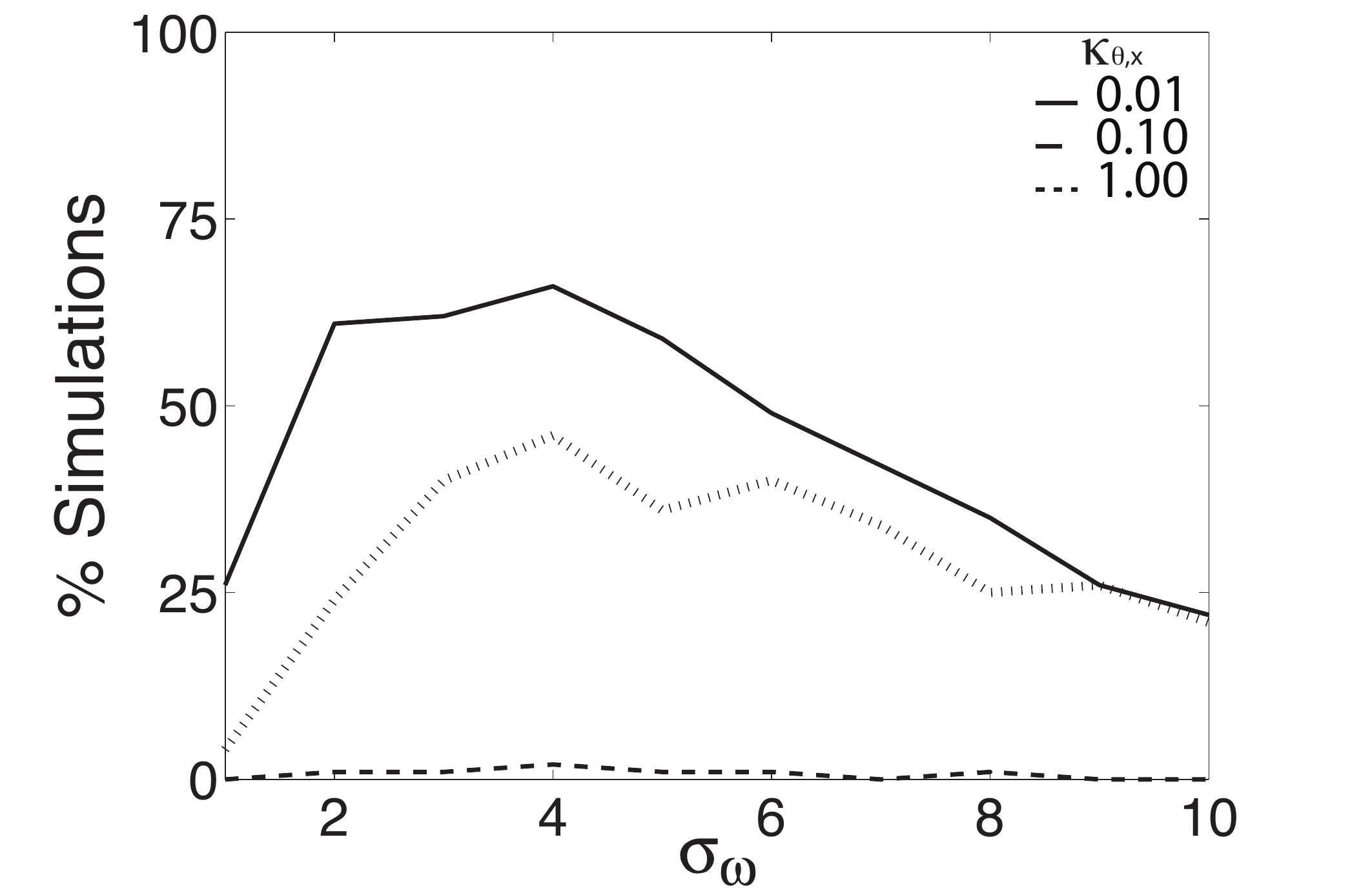}
\label{fig:modelBehaviorSensOsc}
}
\end{center}
\caption{{\bf Percentage of simulations in which the qualitative dynamics in Figure \ref{fig:coupledDynamics} occur.} In (a) oscillators synchronize and switches are ``on", in (b) oscillators freeze and switches are ``off", and in (c) switches vary with oscillators vs $\sigma_{\omega}$ for $\kappa_{\theta,x}=0.01$ (solid), $\kappa_{\theta,x}=0.1$ (dotted) and $\kappa_{\theta,x}=1$ (dashed).  $\kappa_{x,x} = 1 < \hat{\kappa}_{x,x}$, $\kappa_{x,\theta} = 1.5$, and $\kappa_{\theta,\theta} = 40 > \hat{\kappa}_{\theta,\theta}$.} \label{fig:modelBehaviorSens}
\end{figure}

 An additional consequence of coupling switches and oscillators in a state in which switches vacillate between all ``on" and all ``off" along with the synchronized oscillator frequency (Figure~\ref{fig:coupledDynamicsOsc}).  This oscillatory synchronization occurs when the pure Hopfield model would turn switches ``off" ($\kappa_{x,x} < \hat{\kappa}_{x,x}$), the pure Kuramoto model would induce oscillator synchronization ($\kappa_{\theta,\theta} > \hat{\kappa}_{\theta,\theta}$), and the timing between the oscillators and switches are balanced such that the average period of the coupled oscillators is slightly less than the average decay time of the system of switches.  Figure~\ref{fig:modelBehaviorSensOsc} shows that this balance in switch-oscillator timescales increases with decreasing $\kappa_{\theta,x}$ and depends non-monotonically on $\sigma_{\omega}$. As we see in the plot of $r_{\omega}\left(t\right)$ in Figure~\ref{fig:coupledDynamicsOsc}, the average oscillator natural frequencies will decrease towards the end of the ``down" phase in response to switches turning off, and then increase to their full natural values in the ``up" phase as switches turn back ``on".  Therefore, if synchronized oscillator period is too slow (i.e., $\sigma_{\omega}$ is too large), the system will tend to be locked in the ``on" state (Figure~\ref{fig:modelBehaviorSensOn}); if too fast (i.e., $\sigma_{\omega}$ too small) the system will tend to be locked in the ``off" state (Figure~\ref{fig:modelBehaviorSensOff}).  

\subsubsection*{Synchronization of network-wide oscillations may be transitory.}

\par Oscillatory behavior in the switches is also observed for unsynchronized oscillators ($\kappa_{\theta,\theta} < \hat{\kappa}_{\theta,\theta}$) as depicted in Figure~\ref{fig:coupledDynamicsOscNoSyncSwitchOsc}.  In this case, the value of $\kappa_{\theta,x}$ must be large enough to enable switches to freeze the oscillators' phases.  However, because the oscillators are uncoupled, a small subset of oscillators in the ``up" phase can drive the switches to turn on for large-enough values of $\kappa_{x,\theta}$.  These switch oscillations are transitory, ending when at last the switch coupling dominates the system and induces all of the oscillators to freeze.  For unsynchronized oscillators in the parameter range of Figure~\ref{fig:coupledDynamicsOscNoSyncSwitchOsc}, the transitional oscillations in the switch state occurs regularly in 21 of 100 simulations.  In 8 of these 21 simulations, the switch state turns ``on" after decaying at least twice.  More rarely, transitory changes in switch state may be induced by a similar mechanism in simulations for which $\kappa_{\theta,\theta} > \hat{\kappa}_{\theta,\theta}$ and switches ultimately settle on the all ``on" or all ``off" states. 

\subsubsection*{System size affects the distribution of qualitative dynamics}

\par We also explored the dynamics of the coupled system for networks of sizes ranging from $N=M=10$ to $N=M=500$ nodes, described in the methods.  For networks of all sizes, we observe that the dynamics of the system was limited to the four qualitative behaviors observed for networks of size $N=M=100$ depicted in Figure~\ref{fig:coupledDynamics}.  However, the system size does have a notable effect on the frequency with which each of these behaviors occurs.  Supplemental Figures~S5-S7 plot the observed frequencies for each of the network sizes as a function of the $\kappa_{\theta,x}$ and $\sigma_{\omega}$ values considered in Figure~\ref{fig:modelBehaviorSens}. 

\par
\begin{figure}[ht]
\begin{center}
\includegraphics[width=\textwidth]{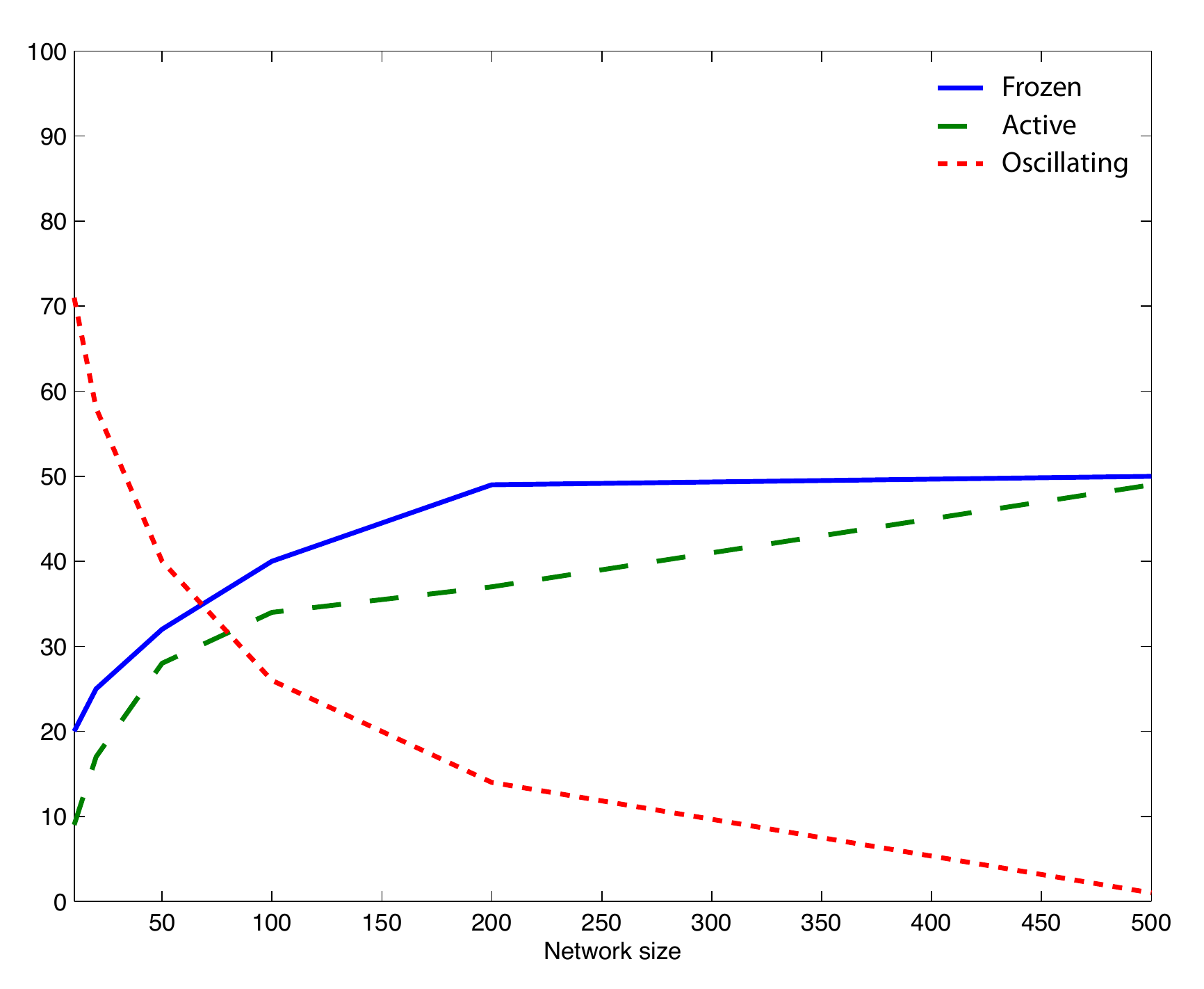}
\end{center}
\caption{\textbf{Dependence on network size for qualitative states for $\kappa_{\theta,x}=0.01$ and $\sigma_{\omega}=1$.} Percentage of simulations in which the qualitative dynamics have switches off and oscillators frozen (blue, solid), switches on and oscillators synchronized (green, dashed), and oscillatory switches and synchronized oscillators (red, dotted).}\label{fig:KMinMinSNetSize}
\end{figure}

 When $\kappa_{\theta,x}=0.01$, the observed frequencies of the system states depend most strongly on network size in simulations using the smallest value of $\sigma_{\omega}=1$ is also small (Supplemental Figure~S5).  In this case, the probability of observing the system with synchronized oscillatory dynamics in both switches and oscillators decays as the network grows.  Both the state in which the switches are on and oscillators are synchronized and the state in which the switches are off and oscillators are frozen have with compensatory increases in probability (Figure~\ref{fig:KMinMinSNetSize}).  The relative probability of obtaining the frozen state increases, with notable decay in the probability of obtaining the state in which switches are ``on'' and oscillators are synchronized in large networks. 

\par
\begin{figure}[ht]
\begin{center}
\includegraphics[width=\textwidth]{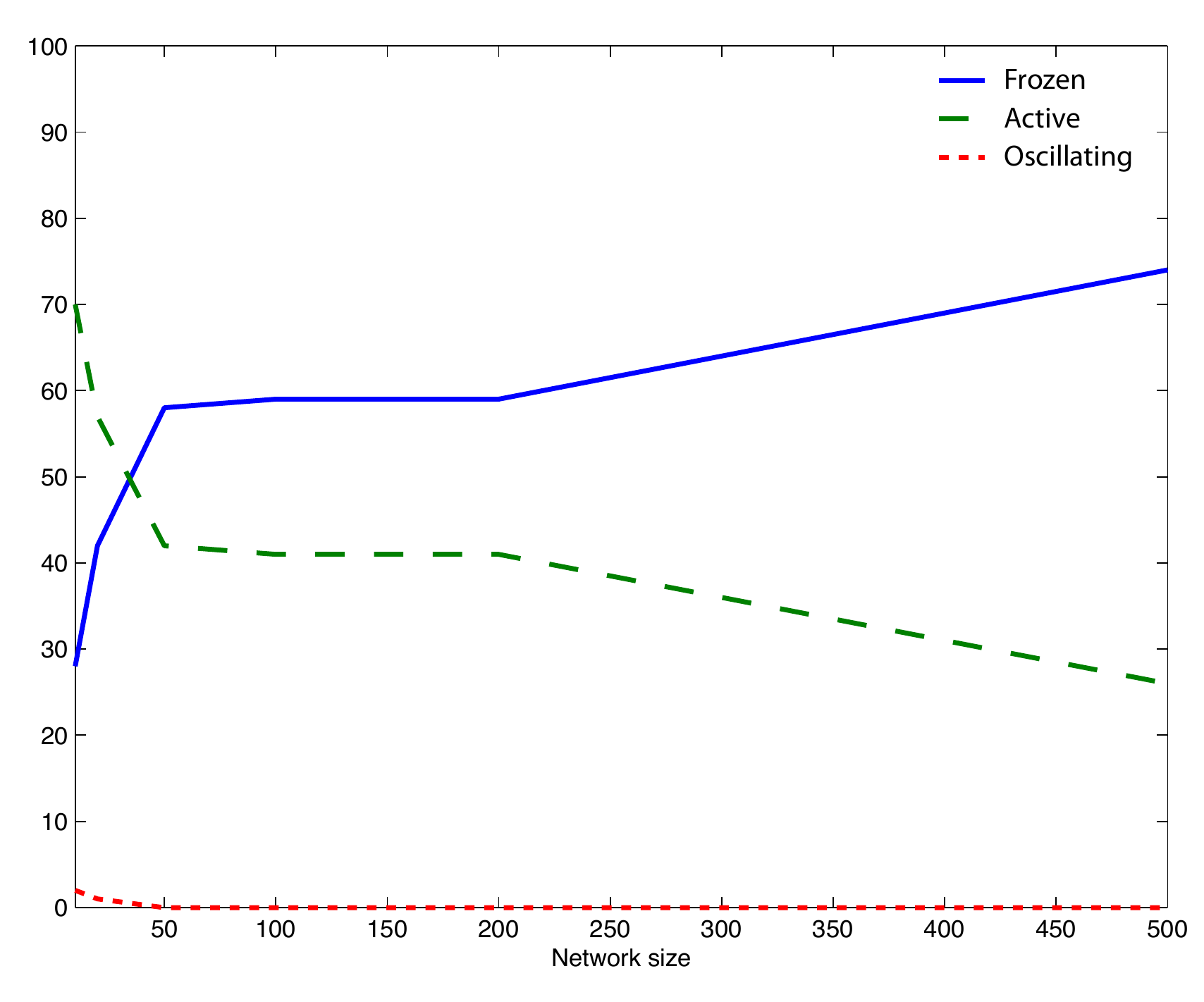}
\end{center}
\caption{\textbf{Dependence on network size for qualitative states for $\kappa_{\theta,x}=1$ and $\sigma_{\omega}=10$.} Percentage of simulations with qualitative dynamics plotted as described in Figure~\ref{fig:KMinMinSNetSize}.}\label{fig:KMaxMaxSNetSize}
\end{figure}

 On the other hand, when $\kappa_{\theta,x}=1$, the system size has the greatest influence on the resulting dynamics for large values of $\sigma_{\omega}$  (Supplemental Figure~S7).  In this case, the system changes from containing mostly switches in the on state and synchronized oscillators to switches that are entirely in the ``frozen" state for large network sizes (Figure~\ref{fig:KMaxMaxSNetSize}).  We hypothesize that the system is forced into the frozen state in larger networks because of increased oscillator synchronization in large networks.  Therefore, small networks would have a higher probability of having few oscillators that are unsynchronized and in the ``up" phase  ($\tilde{\theta} =1$), causing the switches to turn ``on'' ($\tilde{x}=1$) due to the structure of eq.~(\ref{eq:switch}) as was discussed previously.   Furthermore, the rare oscillations observed in both switches and oscillators  when $\kappa_{\theta,x}=1$ occur only when the network is small.  Intermediate values of $\kappa_{\theta,x}=0.1$ show similar changes to those described for $\kappa_{\theta,x}=0.01$ when $\sigma_{\omega}=1$ and to those described for $\kappa_{\theta,x}=1$ when $\sigma_{\omega}=10$ (Supplemental Figure~S6).

\subsection*{The heterogeneous network models qualitative dynamics of the yeast cell cycle derived from network motifs.}

\par
\begin{figure}[ht]
\begin{center}
\includegraphics[width=\textwidth]{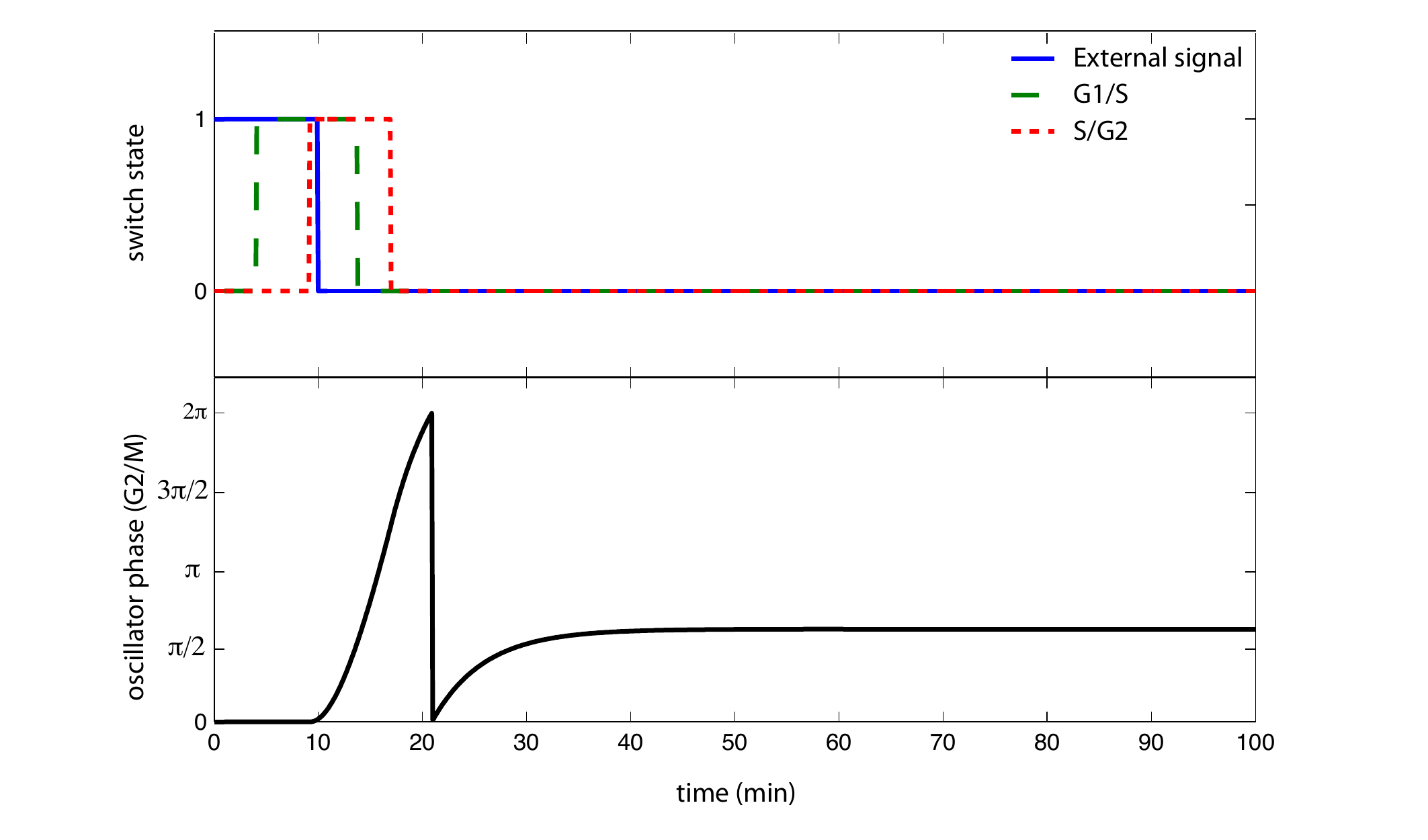}
\end{center}
\caption{\textbf{Simulated dynamics of the yeast cell cycle} Evolution of the states of the cell cycle modules (G1/S top, green dashed; S/G2 top, red dotted; G2/M bottom, black) in response to an external stimulus to initiate the cell cycle (top, blue solid)} \label{fig:normCellCycle}
\end{figure}

\begin{figure}[ht]
\begin{center}
\includegraphics[width=\textwidth]{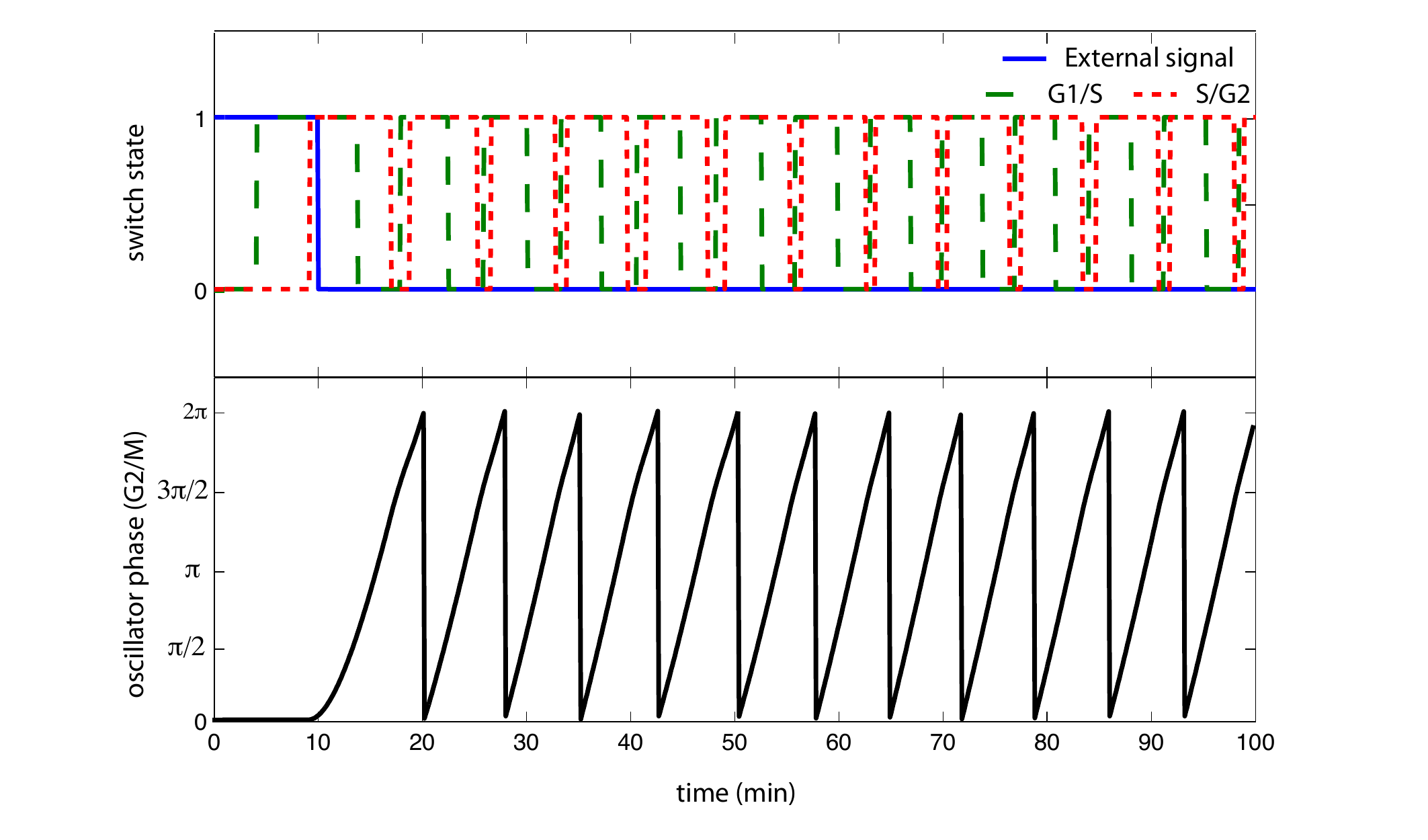}
\end{center}
\caption{\textbf{Simulated dynamics of an aberrant cell cycle network.} As for Figure~\ref{fig:normCellCycle} with a network topology linking the G2/M module to the G1/S module.} \label{fig:cancerCellCycle}
\end{figure}

 Previous work by \cite{Tyson2001} make the cell cycle processes controlling mitotic division of fission yeast \textit{Schizosaccharomyces pombe} cells provides an optimal system in which to apply our model.  The biochemical reactions responsible for driving the cell cycle are well understood and the resulting dynamics in each of the stages of the cell cycle have been characterized extensively in \cite{Tyson2001,Tyson2003,Csikasz-Nagy2008}.  The cell cycle machinery in mitosis is divided into four, sequential stages: phase 1 is a gap or rest phase (G1); phase 2 is a DNA synthesis stage (S); phase 3 is an additional gap stage (G2); and phase 4 is the mitotic division stage (M). Previously, \cite{Tyson2001} observed that the dynamics of the yeast cell cycle can be divided into three sequentially interacting modules, triggered by a signal based upon cell size: (1) G1/S transitions with a toggle-switch, (2) S/G2 transitions with a toggle-switch, and (3) G2/M transitions with an oscillator.  Although the specific timing differs from \cite{Tyson2001}, we observe similar qualitative dynamics to that observed in \cite{Tyson2001} when applying our heterogeneous model to evolve the state of these cell cycle stages (Figure~\ref{fig:normCellCycle}) as described in the Methods section.  We note that the response in this system is consistent with the transitory oscillations observed in Figure~\ref{fig:coupledDynamicsOscNoSyncSwitchOsc} in the case of all-to-all coupling.  We also modeled this cell-cycle system in a rewired-network, in which the G2/M transitions feedback into G1/S (Figure~\ref{fig:cancerCellCycle}).  In this case, we observe sustained reactivation of the cell cycle regardless of the external signal.  These dynamics are analogous to the synchronized dynamics in Figure~\ref{fig:modelBehaviorSensOsc} and consistent with cell growth arising from re-wiring biochemical reactions in cancer cells \cite{Hanahan2000}.

\section*{Discussion}

\par Our model of coupled switches and oscillators in all-to-all networks demonstrates
that networks with components having heterogeneous dynamics can exhibit synchronization similar to that observed in homogeneous systems.  As is the case in homogeneous models (e.g., \cite{Restrepo2005, Restrepo2005a, Restrepo2006,  Restrepo2006a}), we expect analogous synchronization to hold in small-world, biochemical network topologies (e.g., \cite{Barabasi2004}).   However, these alternative topologies would likely change the probability of observing each of the qualitative model behaviors similar to the observed dependence of probabilities in network size.  In this alternative network topologies, the qualitative states of the network model may have greater variability in small network sizes in accordance with the findings of \cite{Pikovsky2002}.  Finally, in these topologies the heterogeneous model could yield additional, complex qualitative dynamic states, resulting from the complex dynamics that they cause in models of coupled switches alone \cite{Edwards2000}.

\par While uncoupled network motifs may adopt switch-like or oscillatory dynamics, coupling between these components can induce switch-like behavior in oscillators and oscillatory behavior in switches.  These qualitative changes in component dynamics occur stochastically, depending on the distribution of frequencies and switch states. They are more likely to occur in simulations with an imbalance in
relative timescales, in which the dynamics of the faster network motif
will dominate the system.  Similarly, when $\kappa_{\theta,x}$ and $\sigma_{\omega}$ are both small, the coordinated oscillations in the switches and oscillators that occur in frequently small networks are largely eliminated in larger networks.  We hypothesize that this larger network effectively increases the range of natural frequencies and phases, making the simulation less likely to have the constrained distribution required to obtain such synchronized oscillations.   We can expect that biological systems have evolved components according to these distributions to ensure the robustness of the dynamics in the system.  For example, multiple proteins can often serve similar functions in cell signaling pathways, which would increase the system size and decrease the probability of transient behaviors in our model.  This robustness will be further ensured through the sheer size of most biochemical systems.  For example, in humans yeast two-hybrid maps and metabolic network maps both contain on the order of thousands of interactions between thousands of species \cite{Barabasi2004}.

\par Furthermore, we have also observed that the heterogeneous network model will
freeze the oscillator dynamics in the presence of inactive switches
and then subsequently activate in synchrony in the presence of active
switches.  As a result, our model provides a natural mechanism for the
coordination of complex machinery such as the initiation of
cell-cycle dynamics.  For example, when we apply our model to the yeast cell cycle motifs in \cite{Tyson2001}, we recapitulate the qualitative dynamics of delayed initiation of stages of the cell cycle observed in simulations with differential equations of the regulatory dynamics in \cite{Tyson2001}.  Additional tuning of the model parameters or incorporation of additional cell cycle checkpoints would facilitate a precise match of the timing of \cite{Tyson2001}.  Because parameters are defined for modules and their interaction, our model requires far fewer rate parameters than any differential equation model of sets of biochemical reactions of the yeast cell cycle.  Generally, the oscillator in the final G2/M step of the cell cycle is active only when the series of switches in the previous steps of the cell cycle are activated, consistent with the transient dynamics observed in our network model.  However, rewiring the network to introduce feedback from the G2/M stage to the G1/S stage of the cell cycle will cause the modeled cell cycle machinery to engage continually without regard to the external growth signals, consistent with the malignant rewring in cancer cells \cite{Hanahan2000}.  Similar to the oscillatory behavior induced in switches in simulations in all-to-all networks, this small modification to the topology of cell cycle interactions altered the resulting dynamics of the network motifs for the G1/S and S/G2 motifs.  We, therefore, hypothesize that motif dynamics predicted by the structure of subgraphs may not accurately describe their \textit{in vivo} dynamics if considered in isolation, consistent with the hypothesis in \cite{Dorogovtsev2010} and findings of \cite{Taylor2011}.

\par We observed that the switches in the cell cycle block activation of the yeast cell cycle when no external signal is present.  Similarly, when part of the larger but sparse networks that compose biochemical systems \cite{Barabasi2004}, inactive switches would effectively destroy links between nodes on the network.  As a result, the proposed heterogeneous model provides a potential mechanism for Kuramoto-based models with evolving network topologies such as \cite{Cumin2007,Mondal2008,Sorrentino2008,Jiang2009,Volman2010,Adhikari2011,Gomez-Gardenes2011}.  Similarly, we observed that the intermediate switches delay the oscillations in the final G2/M motif in the simulated yeast cell cycle.  As a result, we hypothesize that coupling switches to oscillators through their frequencies in this model also provides a natural mechanism for extensions of the Kuramoto model with dynamic frequencies \cite{Cumin2007,Martens2009, Taylor2010} or phase delays  \cite{Wang2008,Ermentrout2009,Wang2009,Wang2011}.

\par The heterogeneous network model described in this
paper facilitates characterization of the dynamics of complex,
biochemical systems by abstracting the dynamics of their composite
motifs such as the yeast cell cycle based upon \cite{Tyson2001}.  We note that the proposed heterogeneous network model is deterministic once the initial values of all the switches and oscillator frequencies have been specified.  However, many intracellular reactions (e.g., \cite{Kaern2005}) and neuronal systems (reviewed in \cite{Stein2005,Faisal2008}) evolve stochastically.  In these cases, the Hopfield networks used to model the switches could be replaced with probabilistic Boolean networks \cite{Shmulevich2002b} and the oscillators evolved with stochastic solvers such as the stochastic simulation algorithm (reviewed in \cite{Gillespie2007}), integrated with the methodology developed in \cite{Fertig2011}.  Similar modifications could also extend the heterogeneous model to incorporate coupling with components of additional dynamics pertinent to biochemical systems, such as those of the network motifs enumerated in \cite{Tyson2003, Alon2007,Novak2008}.  These studies would also ideally consider the dynamics of the heterogeneous network model in additional small-world and random network topologies, as well as the topologies defined by neuronal systems and gene regulatory networks.

\section*{Materials and Methods}

\subsection*{Numerical simulations in the all-to-all network}

\par In this study, we analyze the range of possible
dynamics of the coupled, heterogeneous networks by applying this model
to all-to-all networks.  Analyses were performed for networks with equal number of switches and oscillators ($N=M$) of sizes 10, 50, 100, 200, and 500.   All simulations are run one hundred times from random initial conditions for the state of switches ($x_{i}, i=1,\ldots,N$) and oscillators ($\theta_{j}, j=1,\ldots,M$), drawn from a Gaussian distribution and a uniform distribution on $[0,2\pi)$, respectively.  Similarly, oscillator natural frequencies are drawn randomly from a Gaussian distribution of mean zero and standard deviation parameter $\sigma_{\omega}$.  Simulations of 100~seconds (in the arbitrary units of the model), with a time step of 0.01~seconds were found sufficient to reflect the range of possible model behaviors and verify consistency across initial conditions.  The behavior of each simulation is summarized based on the time-dependent order parameters $r_{\theta}\left(t\right)$ and $\psi\left(t\right)$, $r_{\omega}\left(t\right)$, and $r_{x}\left(t\right)$. 

\subsection*{Numerical simulations of the yeast cell cycle}

\par Based upon \cite{Tyson2001}, we model the yeast cell cycle as an initiating external signal (namely the cell size), coupled to a toggle switch representing the transition between G1/S, a toggle switch representing the transition between S/G2, and an oscillator representing the transition from G2/M.  While the external signal is incorporated into the model with coupling to the other switches in eq.~(\ref{eq:genSwitch}), its state is not updated by the model.  The duration of this external signal is set at 10 simulated minutes, based upon \cite{Tyson2001}.  Similarly, the initial values of the hidden variable $x$ for the switches in the G1/S and S/G2 modules are set at -0.5, $\tau$ to 1, and $\kappa_{x,x}$ to 2 to reproduce the approximate 10 minute duration of these switches in \cite{Tyson2001}.   The natural frequency is for the G2/M module set to $\frac{2\pi}{60} \mbox{min}^{-1}$ to likewise reflect the timescale reported in \cite{Tyson2001}, while the remainder of the coupling parameters are left untuned, set to $\kappa_{\theta,x} = \kappa_{x,\theta} = \kappa_{\theta,\theta}=2$ because we sought only to reproduce the qualitative dynamics of the \cite{Tyson2001} model.  The rewiring in the system with enduring cell cycle activation is achieved by adding an edge from the module for G2/M to the switch in the G1/S module.

\section*{Acknowledgments}

This work was sponsored by NCI (CA141053).  We thank LV Danilova, B Fertig, AV Favorov, BR Hunt, MJ Stern, MF Ochs, E Ott, E Webster, and LM Weiner for advice.  Code available upon request.

\bibliography{correlations}

\newpage

\section*{Figure Legends}

\listoffigures

\newpage

\newpage

\subsection*{Supplemental Figure Captions}

\par \textbf{Figure S5.  Dependence of dynamics on network size for $\kappa_{\theta,x}=0.01$.}   Number of simulations (of 100) for which the switches are off and oscillators are frozen (left panel), the switches are on and the oscillators are synchronized (center panel), and both the oscillators and switches have synchronized oscillations (right).

\par \textbf{Figure S6. Dependence of dynamics on network size for $\kappa_{\theta,x}=0.1$.} As for Supplemental Figure~S5.

\par \textbf{Figure S7. Dependence of dynamics on network size for $\kappa_{\theta,x}=1$.} As for Supplemental Figure~S5.

\end{document}